\documentclass[aps,showpacs,preprintnumbers,amsmath,amssymb]{revtex4}
\usepackage{dcolumn}
\usepackage[dvips]{epsfig}
\usepackage{graphicx}
\usepackage{amssymb}
\usepackage{color}
\usepackage{enumerate}

\begin{document}
\baselineskip=0.8 cm
\title{\bf Gravitational field of a slowly rotating black hole with phantom global monopole}

\author{Songbai Chen\footnote{csb3752@hunnu.edu.cn}, Jiliang Jing
\footnote{jljing@hunnu.edu.cn}}

\affiliation{Institute of Physics and Department of Physics, Hunan
Normal University,  Changsha, Hunan 410081, People's Republic of
China \\ Key Laboratory of Low Dimensional Quantum Structures \\
and Quantum Control of Ministry of Education, Hunan Normal
University, Changsha, Hunan 410081, People's Republic of China}

\begin{abstract}
\baselineskip=0.6 cm
\begin{center}
{\bf Abstract}
\end{center}

We present a slowly rotating black hole with phantom global monopole
by solving Einstein's field equation and find that presence of
global monopole changes the structure of black hole. The metric
coefficient  $g_{t\phi}$ contains hypergeometric function of the
polar coordinate $r$, which is more complex than that in the usual
slowly rotating black hole. The energy scale of symmetry breaking
$\eta$ affects the black hole horizon and a deficit solid angle.
Especially, the solid angle is surplus rather than deficit for a
black hole with the phantom global monopole. We also study the
correction originating from the global monopole to the angular
velocity of the horizon $\Omega_H$, the Kepler's third law, the
innermost stable circular orbit and the radiative efficiency
$\epsilon$ in the thin accretion disk model. Our results also show
that for the phantom black hole the radiative efficiency $\epsilon$
is positive only for the case $\eta\leq \eta_c$. The threshold value
$\eta_c$ increases with the rotation parameter $a$.

\end{abstract}

 \maketitle
\newpage
\section{Introduction}

Phantom field is a special kind of dark energy model with the
negative kinetic energy \cite{Caldwell}, which is applied
extensively in cosmology to explain the accelerating expansion of
the current Universe \cite{ph1,ph2,ph3,ph4,ph5,ph6}. Comparing with
other dark energy models, the phantom field is more interesting
because that the presence of negative kinetic energy results in that
the equation of state of the phantom field is less than $-1$ and
then the null energy condition is violated.  Although the phantom
field owns such exotic properties, it is still not excluded by
recent precise observational data \cite{ph7}, which encourages many
people to focus on investigating phantom field from various aspects
of physics.

Phantom field also exhibits some peculiar properties in the black
hole physics. E. Babichev \cite{EBab} found that the mass of a black
hole decreases when it absorbs the phantom dark energy. This means
that the cosmic censorship conjecture is challenged severely by a
fact that the charge of a Reissner-Nordstr\"{o}m-like black hole
absorbing the phantom energy will be larger than its mass.  We
studied the wave dynamics of the phantom scalar perturbation in the
Schwarzschild black hole spacetime and that in the late-time
evolution the phantom scalar perturbation grows with an exponential
rate rather than decays as the usual scalar perturbations
\cite{Sb,bw}. Moreover, we also find that the phantom scalar
emission will enhance the Hawking radiation of a black hole
\cite{Sb1}. Furthermore, some black hole solutions describing
gravity coupled to phantom scalar fields or phantom Maxwell fields
have also been found in
\cite{pbh1,pbh2,pbh3,pbh4,pbh5,pbh6,pbh7,pbh8}. The thermodynamics
and the possibility of phase transitions in these phantom black
holes are studied in \cite{pbh4,pbh5}. The gravitational collapse of
a charged scalar field \cite{pbh6} and the light paths \cite{pbh7}
are investigated in such kind of spacetimes. Moreover, S. Bolokhov
\textit{et al} also study the regular electrically and magnetically
charged black hole with a phantom scalar in \cite{pbh8}. These
investigations could help us to get a deeper understand about dark
energy and black hole physics.

A global monopole is one of the topological defects which could be
formed during phase transitions in the evolution of the early
Universe. The metric describing a static black hole with a global
monopole was obtained by Barriola and Vilenkin \cite{Barriola},
which arises from the breaking of global SO(3) symmetry of a triplet
scalar field in a Schwarzschild background. Due to the presence of
the global monopole, the black hole owns different topological
structure from that of the Schwarzschild black hole. The physical
properties of the black hole with a global monopole have been
studied extensively in recent years \cite{gb1,gb2,gb3,gb4}.

The main purpose of this paper is to study the gravitational field
of phantom global monopole arising from a triplet scalar field with
negative kinetic energy and to see how the energy scale of symmetry
breaking $\eta$ influences the structure of black hole, the Kepler's
third law, the innermost stable circular orbit and the radiative
efficiency $\epsilon$ in the thin accretion disk model. Moreover, we
will explore how it differs from that in the black hole with
ordinary global monopole.

The paper is organized as follows: in the following section we will
construct a static and spherical symmetric solution of a phantom
global monopole from a triplet scalar field with negative kinetic
energy, and then study the effect of the parameter $\eta$ on the
black hole. In Sec.III, we obtain a slowly rotating black hole with
phantom global monopole by solving Einstein's field equation and
find that presence of global monopole make  the metric coefficient
of black hole more complex. In Sec.IV, we will focus on
investigating the effects of the parameter $\eta$ on the Kepler's
third law, the innermost stable circular orbit and the radiative
efficiency $\epsilon$ in the thin accretion disk model. We end the
paper with a summary.

\section{A static and spherical symmetric black hole with phantom global monopole}

Let us now first study a static and spherical symmetric black hole
with phantom global monopole formed by spontaneous symmetry breaking
of a triplet of phantom scalar fields with a global symmetry group
$O(3)$. The action giving rise to the phantom global monopole is
\begin{eqnarray}
S=\int \sqrt{-g}d^4x[R -\frac{\xi}{2}\partial^{\mu}\psi^a\partial_{\mu}\psi^i-\frac{\lambda}{4}
(\psi^i\psi^i-\eta^2)^2],\label{act1}
\end{eqnarray}
where $\psi^i$ is a triplet of scalar field with $i=1, 2, 3$, $\eta$
is the energy scale of symmetry breaking and $\lambda$ is a
constant. The coupling constant $\xi$ in the kinetic term takes the
value $\xi=1$ corresponds to the case of the ordinary global
monopole originating from the scalar field with the positive kinetic
energy \cite{Barriola}. As the coupling constant $\xi=-1$, the
kinetic energy of the scalar field is negative and then the phantom
global monopole is formed.

Following in Ref.\cite{Barriola}, we can take ansatz describing a monopole as
\begin{eqnarray}
\psi^a=\frac{\eta f(r)x^i}{r},
\end{eqnarray}
where $x^ix^i=r^2$. Equipping with the general static and spherical
symmetric metric
\begin{eqnarray}
ds^2=-B(r)dt^2+A(r)dr^2+r^2d\theta^2+r^2\sin^2{\theta}
d\phi^2, \label{metric0}
\end{eqnarray}
one can find that the field equations for $\psi^i$ can be reduced to a single equation for $f(r)$
\begin{eqnarray}
\frac{\xi
f''}{A}+\bigg[\frac{2}{Ar}+\frac{1}{2B}\bigg(\frac{B}{A}\bigg)'\bigg]\xi
f'-\frac{2\xi f}{r^2}-\lambda \eta^2f(f^2-1)=0.
\end{eqnarray}
Moreover, the energy-momentum tensor for the spacetime with a global monopole can be expressed as
\begin{eqnarray}
T^t_t&=&\bigg[\frac{\eta^2f'^2}{2A}+\frac{\eta^2f^2}{r^2}\bigg]\xi
+\frac{\lambda}{4}\eta^4(f^2-1)^2,\\
T^r_r&=&\bigg[-\frac{\eta^2f'^2}{2A}+\frac{\eta^2f^2}{r^2}\bigg]\xi
+\frac{\lambda}{4}\eta^4(f^2-1)^2,\\
T^{\theta}_{\theta}&=&T^{\phi}_{\phi}=\frac{\xi\eta^2f'^2}{2A}
+\frac{\lambda}{4}\eta^4(f^2-1)^2.
\end{eqnarray}
Similarly, as in Ref.\cite{Barriola}, one can take an approximation
$f(r)=1$ outside the core due to a fact that $f(r)$ grows linearly
when $r <(\eta\sqrt{\lambda})^{-1}$ and tends exponentially to unity
as soon as $r >(\eta\sqrt{\lambda})^{-1}$. With this approximation,
we can obtain a solution of the Einstein equations
\begin{eqnarray}
B=A^{-1}=1-8\pi\xi\eta^2-\frac{2M}{r},\label{fba11}
\end{eqnarray}
where $M$ is an integrate constant. Obviously, the radius of event
horizon is $r_H=2M/(1-8\pi\xi\eta^2)$. Here we must point that it is
possible that the radius of event horizon $r_H$ is larger than the
monopole's core $\delta\sim(\eta\sqrt{\lambda})^{-1}$ if the
coupling constant
$\lambda\gg\frac{(1-8\pi\xi\eta^2)^2}{4M^2\eta^2}$. In this case,
the metric (\ref{metric0}) can describe the geometry near the
horizon in the spacetime with global monopole. With the increase of
the energy scale of symmetry breaking $\eta$, one can find that the
radius of event horizon $r_H$ increases for a Schwarzschild black
hole with the ordinary global monopole ($S_+$), but decreases for a
Schwarzschild black hole with the phantom global monopole ($S_-$).
Thus, comparing with the system $S_+$, one can find that the system
$S_-$ possesses the higher Hawking temperature and the lower
entropy. In the low energy limit, the luminosity of Hawking
radiation of a spherical symmetric black hole can be approximated as
$L=\frac{2\pi^3r^2_H}{15}T^4_H\propto(1-8\pi\xi\eta^2)^4/r^2_H$,
which tells us that the energy scale of symmetry breaking $\eta$
enhances Hawking radiation for the system $S_-$, but it decreases
Hawking radiation in the system $S_+$. The presence of the phantom
field enhances the Hawking emission of Kerr black hole are also
found in \cite{Sb1}.

Introducing the following
transformations
\begin{eqnarray}
t\rightarrow (1-8\pi\xi\eta^2)^{1/2}t,~~~~~~r\rightarrow
(1-8\pi\xi\eta^2)^{1/2}r,~~~~~M\rightarrow(1-8\pi\xi\eta^2)^{3/2}M,\label{bhua1}
\end{eqnarray}
one can rewrite the metric (\ref{metric0}) with the functions (\ref{fba11}) as
\begin{eqnarray}
ds^2&=&-\bigg(1-\frac{2M}{r}\bigg)dt^2+\bigg(1-\frac{2M}{r}\bigg)^{-1}dr^2+
(1-8\pi\xi\eta^2)r^2(d\theta^2+\sin^2{\theta}d\phi^2).
\label{metric1}
\end{eqnarray}
It is clear that there exists a deficit solid angle $(1-8\pi
\eta^2)$ for the system $S_+$. However, for the system $S_-$ (i.e.,
$\xi=-1$),  one can find that the solid angle becomes $(1+8\pi
\eta^2)$, which is surplus rather than deficit. This implies that
the topological properties of a spacetime with the phantom global
monopole is different from that of with a ordinary global monopole.

\section{A slowly rotating black hole with phantom global monopole}

In this section, we first obtain the metric for a slowly rotating
black hole with phantom global monopole by solving Einstein's field
equation.  And then, we will study the properties of the black hole
spacetime.

From the static and spherical symmetric solution (\ref{metric0})
with the metric function (\ref{fba11}), we can assume the metric has
a form for a slowly rotating black hole with phantom global monopole
\begin{eqnarray}
ds^2&=&-U(r)dt^2+\frac{1}{U(r)}dr^2-2F(r,\theta)adtd\phi+r^2(d\theta^2+\sin^2{\theta}
d\phi^2), \label{metric3}
\end{eqnarray}
where $a$ is a parameter associated with its angular momentum. Moreover, we assume
that in the slowly rotating spacetime (\ref{metric3}) the ansatz describing a monopole is
\begin{eqnarray}
\psi^a=\frac{\eta f(r)x^i}{r}.\label{gms1}
\end{eqnarray}
It is easy to find that the field equations for $\psi^i$ can also be reduced to a single equation for $f(r)$
\begin{eqnarray}
\xi f''U(r)+\bigg[\frac{2U(r)}{r}+U(r)'\bigg]\xi f'-\frac{2\xi f}{r^2}-\lambda \eta^2f(f^2-1)=0,
\end{eqnarray}
which is similar to that in the static and spherical symmetric
spacetime (\ref{metric0}). It is not surprising since the triplet
scalar field $\psi^{i}$ does not depend on the time coordinate $t$.

Inserting the metric (\ref{metric3}) and the triplet scalar
(\ref{gms1}) into the Einstein field equation, we find that the
non-vanishing components of field equation can be expanded to first
order in the angular momentum parameter $a$ as
\begin{eqnarray}
tt: &&\frac{U(r)}{r^2}\bigg[U'(r)r+U(r)-1\bigg]
+8\pi U(r)\bigg[\bigg(\frac{\eta^2f'^2U(r)}{2}+\frac{\eta^2f^2}{r^2}\bigg)\xi
+\frac{\lambda}{4}\eta^4(f^2-1)^2\bigg]=0+\mathcal{O}(a^2),\label{gt1}\\
rr:
&&\frac{1}{U(r)r^2}\bigg[U'(r)r+U(r)-1\bigg]+
\frac{8\pi}{U(r)}\bigg[\bigg(-\frac{\eta^2f'^2U(r)}{2}+\frac{\eta^2f^2}{r^2}\bigg)\xi
+\frac{\lambda}{4}\eta^4(f^2-1)^2\bigg]=0+\mathcal{O}(a^2)\label{gt2},\\
\theta \theta:&&-\frac{r}{2}\bigg[U''(r)r+2U'(r)\bigg]+8\pi r^2\bigg[\frac{\xi\eta^2f'^2U(r)}{2}
+\frac{\lambda}{4}\eta^4(f^2-1)^2\bigg]=0+\mathcal{O}(a^2)\label{gt3},\\
\phi\phi:&&-\frac{r}{2}\sin^2\theta\bigg[U''(r)r+2U'(r)\bigg]+8\pi r^2\bigg[\frac{\xi\eta^2f'^2U(r)}{2}
+\frac{\lambda}{4}\eta^4(f^2-1)^2\bigg]\sin^2\theta=0+\mathcal{O}(a^2)\label{gt4},\\
t\phi:&&\frac{1}{2r^2}\bigg\{r^2U(r)\frac{\partial^2F(r,\theta)}{\partial
r^2}-F(r,\theta)\bigg[r^2U''(r)+2rU'(r)+2U(r)\bigg]+2F(r,\theta)+
\frac{\partial^2F(r,\theta)}{\partial \theta^2}-\frac{\partial
F(r,\theta)}{\partial \theta}\cot\theta\bigg\}\nonumber\\&&-8\pi
F(r,\theta)\bigg[\bigg(\frac{\eta^2f'^2U(r)}{2}+\frac{\eta^2f^2}{r^2}\bigg)\xi
+\frac{\lambda}{4}\eta^4(f^2-1)^2\bigg]=0+\mathcal{O}(a^2)\label{gt5}.
\end{eqnarray}
Solving the Einstein equations (\ref{gt1})-(\ref{gt4}) with the
approximation $f(r)=1$ outside the core, we can obtain the metric
coefficient
\begin{eqnarray}
U(r)=1-8\pi\xi\eta^2-\frac{2M}{r}.\label{fba22}
\end{eqnarray}
Separating $F(r,\theta)=h(r)\Theta(\theta)$, we can obtain the equation for the angular part
\begin{eqnarray}
\frac{d^2\Theta(\theta)}{d\theta^2}-\frac{d\Theta(\theta)}{d\theta}\cot\theta
=\lambda\Theta(\theta).\label{angu}
\end{eqnarray}
In order to that the coefficient $g_{t\phi}$ can be reduced to
that in the slowly rotating black hole without the global monopole,
here we set $\lambda=-2$ and find that $\Theta(\theta)=\sin^2(\theta)$ in this case.
And then the radial part of Eq.(\ref{gt5}) becomes
\begin{eqnarray}
r^2U(r)\frac{d^2h(r)}{dr^2}-2h(r)U(r)-16\pi\xi\eta^2h(r)=0.\label{radial}
\end{eqnarray}
Substituting Eq. (\ref{fba22}) into the above radial equation, we obtain (see in appendix)
\begin{eqnarray}
h(r)=\bigg[\frac{2M}{(1-b)r}\bigg]^{\frac{1}{2}(\sqrt{\frac{9-b}{1-b}}-1)}
~_2F_1\bigg[\frac{1}{2}\bigg(\sqrt{\frac{9-b}{1-b}}-3\bigg),
\frac{1}{2}\bigg(\sqrt{\frac{9-b}{1-b}}+3\bigg),\sqrt{\frac{9-b}{1-b}}+1,
\frac{2M}{(1-b)r}\bigg],\nonumber\\ \label{hr}
\end{eqnarray}
where $b=8\pi\xi\eta^2$ and $_2F_1[a_1,b_1,c_1;x]$ is the
hypergeometric function. As a usual slowly rotating black hole, the
horizon of black hole (\ref{metric3}) is given by the zeros of the
function $U(r)=(g_{rr})^{-1}$, i.e., $r_H=\frac{2M}{1-b}$, which is
the same as that in the static and spherical symmetric case
(\ref{metric0}). The mainly reason is that we here expand the metric
only to first order in the angular momentum parameter $a$.

From Eq. (\ref{hr}), it is obvious to see that due to the presence
of the global monopole the form of the metric coefficient
$g_{t\phi}$ becomes more complicated in the slowly rotating black
hole (\ref{metric3}). The dependence of $h(r)$ on the parameter
$\eta$ is shown in Fig. (1), which tells us that for the slowly
rotating black hole with the ordinary global monopole ($SR_+$) the
function $h(r)$ increases with the parameter $\eta$ near the
horizon, but decreases at the far field region with the larger value
of $r$. For a slowly rotating black hole with the phantom global
monopole ($SR_-$), the behavior of $h(r)$ is just the opposite. In a
word, the dependence of the function $h(r)$ on the $\eta$ near the
horizon is different that in the far field region in these two
global monopole cases.
\begin{figure}[ht]
\begin{center}
\includegraphics[width=7cm]{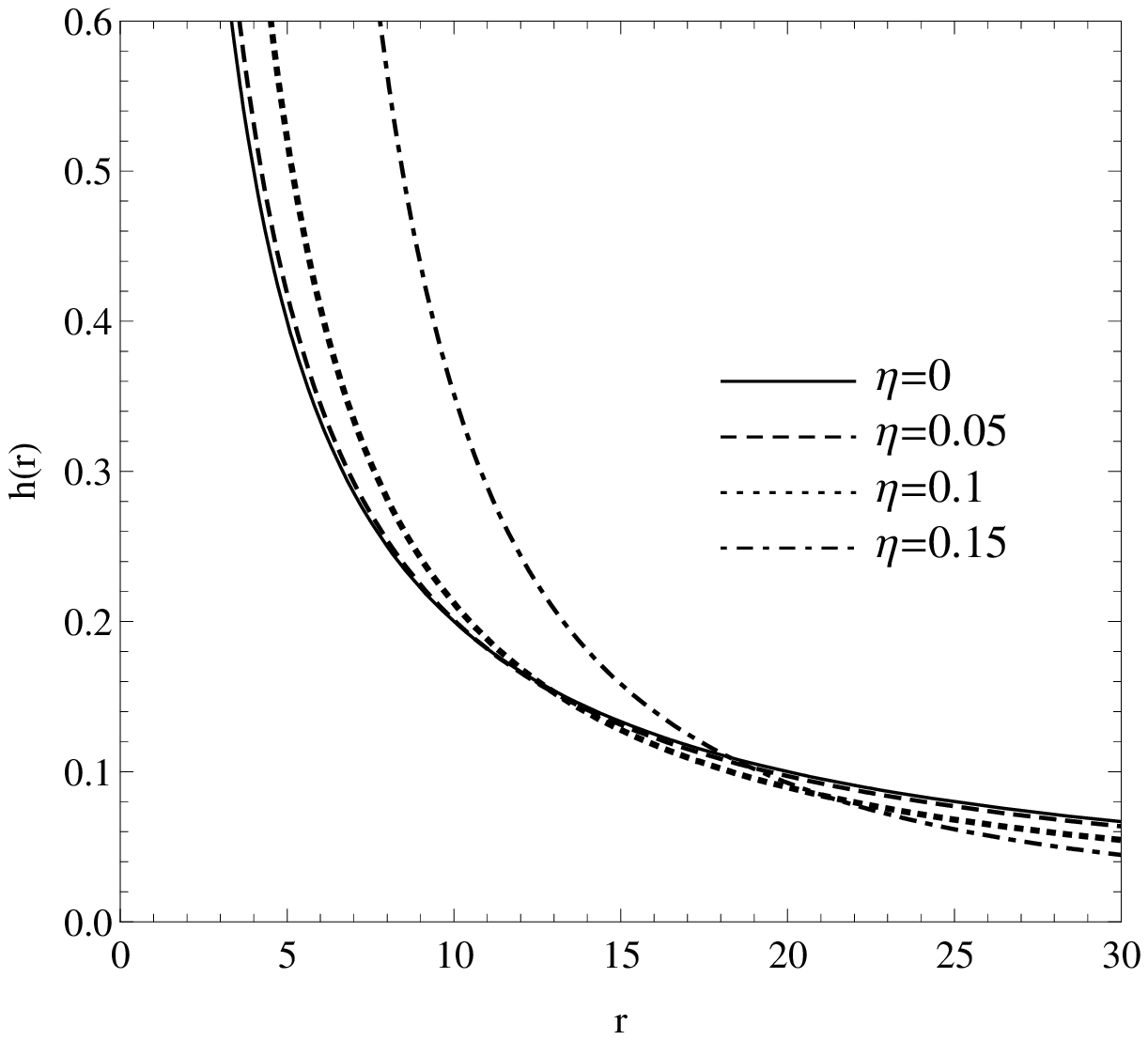}\includegraphics[width=7cm]{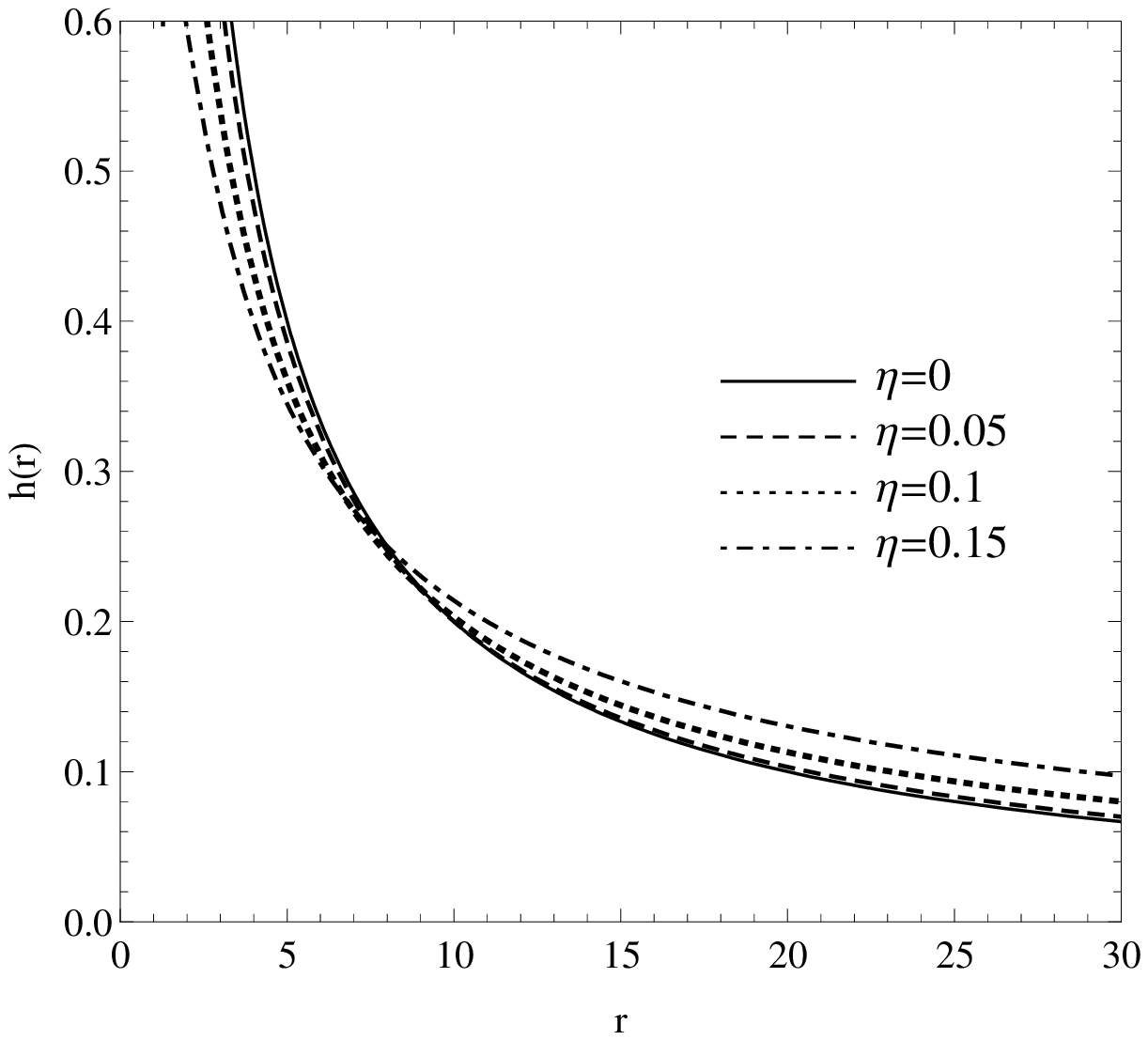}
\caption{The dependence of $h(r)$ on the parameter $\eta$.  The left
and right panels are for the systems $SR_+$ and $SR_-$,
respectively. Here we set $M=1$.}
\end{center}
\end{figure}
As the parameter $b\rightarrow 0$, one can find that
$h(r)\rightarrow \frac{2M}{r}$, which recovers that of a slowly
rotating black hole without global monopole. When the rotation
parameter $a$ vanishes, one can get the previous solution of a
static and spherical symmetric black hole with phantom global
monopole (\ref{metric0}).

For a rotating black hole, one of the important quantities is the
angular velocity of the horizon $\Omega_H$, which affects the region
where  the super-radiance occurs in the black hole background. In
the spacetime of a slowly rotating black hole with global monopole,
the angular velocity of the horizon $\Omega_H$ can be expressed as
\begin{eqnarray}
\Omega_H=-\frac{g_{t\phi}}{g_{\phi\phi}}\bigg|_{r=r_H}
=\frac{a(1-b)^2}{r^2_H}\frac{\Gamma[\sqrt{\frac{9-b}{1-b}}+1]}
{4\Gamma[\frac{1}{2}(\sqrt{\frac{9-b}{1-b}}-1)]
\Gamma[\frac{1}{2}(\sqrt{\frac{9-b}{1-b}}+5)]},
\end{eqnarray}
which depends on the energy scale of symmetry breaking $\eta$. We
plot the change of the angular velocity $\Omega_H$ with the
parameter $\eta$ in Fig.(2). For the system $SR_+$, we find that the
angular velocity $\Omega_H$ first decreases slowly and then
increases rapidly with the increase of $\eta$. There exists a
minimum for $\Omega_H$ at where $\eta=0.1725$ (i.e., $b=0.7477$).
This means that there exist a minimum region for the occurrence of
the super-radiance in the black hole with the ordinary global
monopole for fixed $a$. For the the system $SR_-$,  $\Omega_H$
increases monotonically with the energy scale of symmetry breaking
$\eta$.
\begin{figure}[ht]
\begin{center}
\includegraphics[width=7cm]{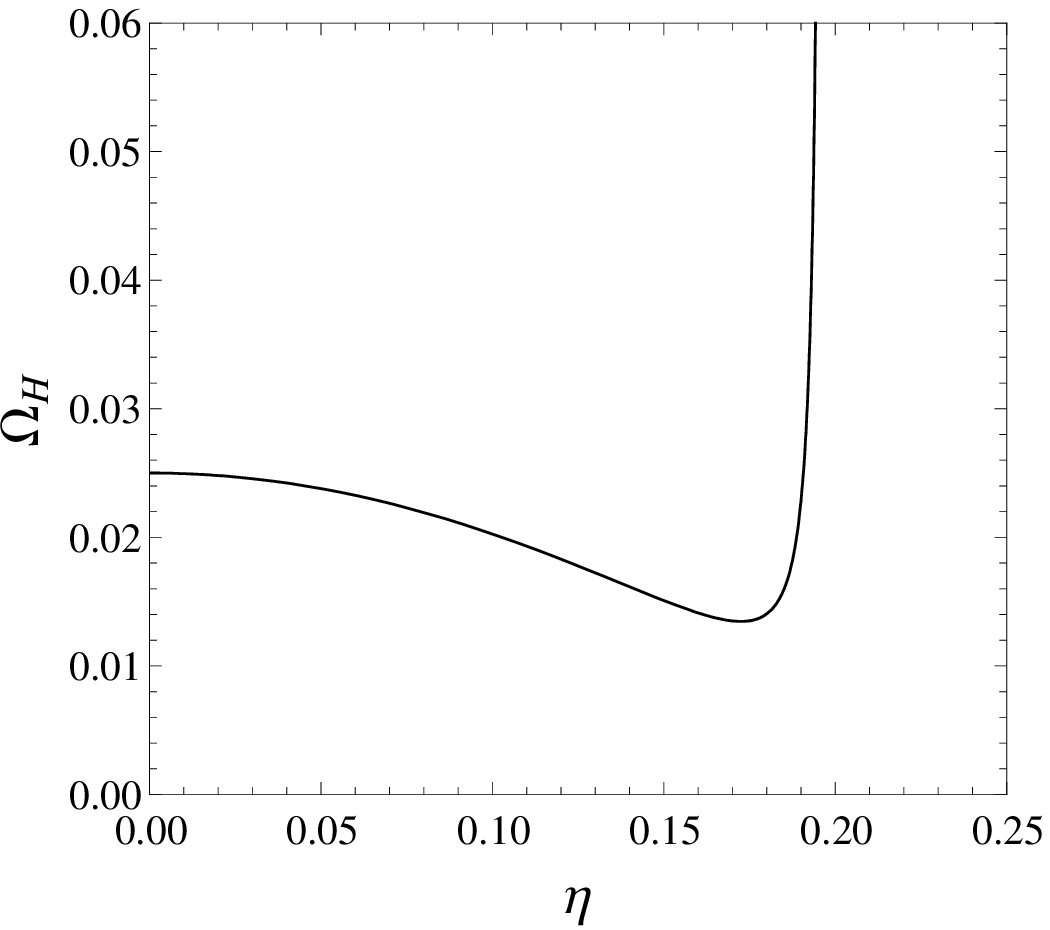}\includegraphics[width=7cm]{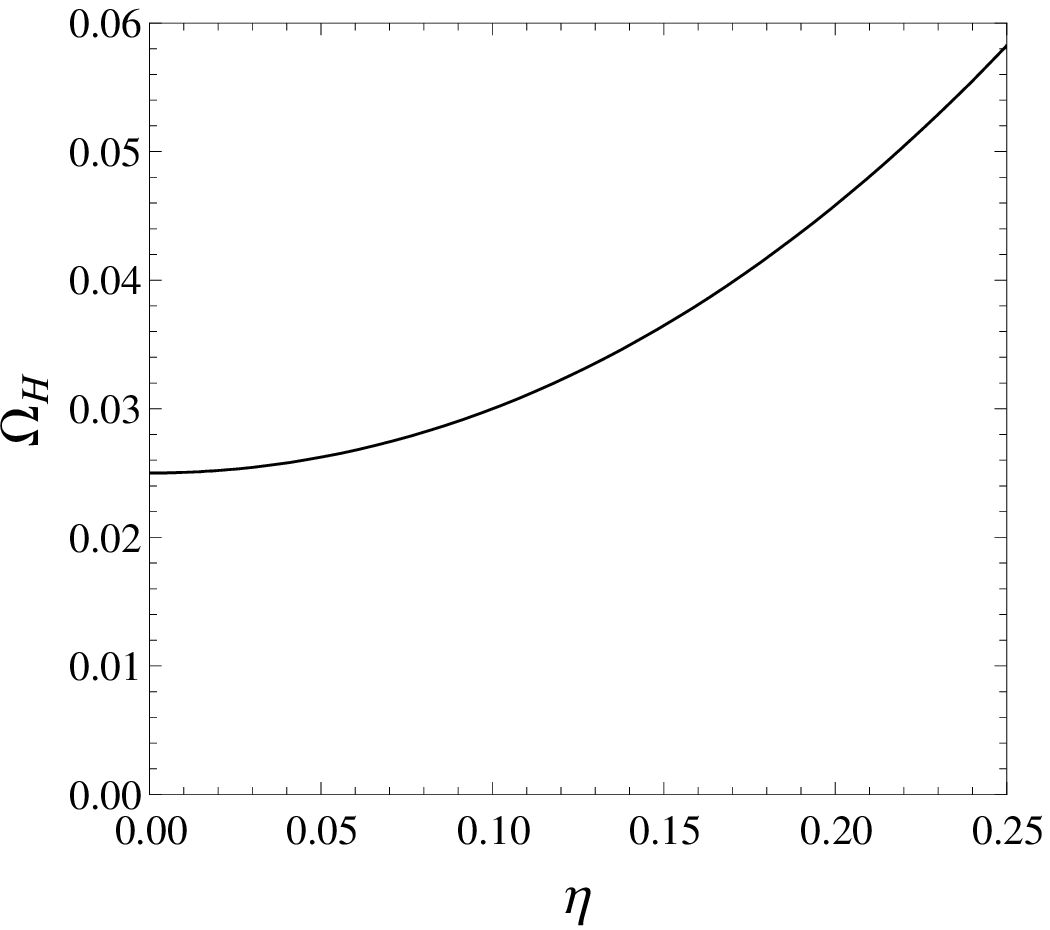}
\caption{The change of the angular velocity $\Omega_H$ with the
parameter $\eta$. The left and right panels are for the systems
$SR_+$ and $SR_-$, respectively. Here we set $M=1$ and $a=0.2$.}
\end{center}
\end{figure}

Making the same transformations (\ref{bhua1}) as in the Sec.II, one
find that the area of the horizon becomes $A_H=br^2_H $. It means
that for the system $SR_-$, the solid angle is surplus rather than
deficit, which is similar to that in the static and spherical
symmetric case (\ref{metric0}).

\section{Kepler's Third Law and the innermost stable circular orbit}

In this section, we will focus on how the symmetry breaking scale
$\eta$ of global monopole affects the Kepler's third law and the
innermost stable circular orbit (ISCO) in this slowly rotating black
hole.

In the stationary and axially symmetric spacetime, one can find that the
timelike geodesics take the form
\begin{eqnarray}
&&\dot{t}=\frac{Eg_{\phi\phi}-L_zg_{t\phi}}{g^2_{t\phi}+g_{tt}g_{\phi\phi}},\label{u1}\\
&&\dot{\phi}=\frac{Eg_{t\phi}+L_zg_{tt}}{g^2_{t\phi}+g_{tt}g_{\phi\phi}},\label{u2}\\
&&g_{rr}\dot{r}^2+g_{\theta\theta}\dot{\theta}^2=V_{eff}(r,\theta; E,L_z),\label{u3}
\end{eqnarray}
with the effective potential
\begin{eqnarray}
V_{eff}(r)=\frac{E^2g_{\phi\phi}-2EL_zg_{t\phi}-L^2_zg_{tt}
}{g^2_{t\phi}+g_{tt}g_{\phi\phi}}-1,
\end{eqnarray}
where the overhead dot stands for a derivative with respect to the
affine parameter. The constants $E$ and $L_z$ correspond to the
conserved energy and the ($z$-component of) orbital angular momentum
of the particle, respectively.

For simplicity, we set the orbits on the equatorial plane. With the
restriction that $\theta=\pi/2$,  one can find that for the stable
circular orbit in the equatorial plane, the effective potential
$V_{eff}(r)$ must obey
\begin{eqnarray}
V_{eff}(r)=0, \;\;\;\;\;\;\;\frac{dV_{eff}(r)}{dr}=0.
\end{eqnarray}
Solving above equations, one can obtain
\begin{eqnarray}
&&E=\frac{g_{tt}+g_{t\phi}\Omega}{\sqrt{g_{tt}+2g_{t\phi}\Omega-g_{\phi\phi}\Omega^2}},\nonumber\\
&&L_z=\frac{-g_{t\phi}+g_{\phi\phi}\Omega}{\sqrt{g_{tt}+2g_{t\phi}\Omega-g_{\phi\phi}\Omega^2}},\nonumber\\
&&\Omega=\frac{d\phi}{dt}=\frac{g_{t\phi,r}+\sqrt{(g_{t\phi,r})^2
+g_{tt,r}g_{\phi\phi,r}}}{g_{\phi\phi,r}},\label{jsd}
\end{eqnarray}
where $\Omega$ is the angular velocity of particle moving in the
orbits. From Eq.(\ref{jsd}), one can obtain Kepler's third law in
the slowly-rotating black-hole spacetime with the global monopole
\begin{eqnarray}
T^2&=&\frac{4\pi^2R^3}{M}\bigg[1+\frac{4a}{(1-b)^2(\sqrt{\frac{9-b}{1-b}}+1)M^{1/2}R^{3/2}}
\bigg[\frac{2M}{(1-b)R}\bigg]^{\frac{1}{2}(\sqrt{\frac{9-b}{1-b}}-1)}
\bigg\{(1-b)R\times\nonumber\\&&~_2F_1\bigg[\frac{1}{2}\bigg(\sqrt{\frac{9-b}{1-b}}-3\bigg),
 \frac{1}{2}\bigg(\sqrt{\frac{9-b}{1-b}}+3\bigg),\sqrt{\frac{9-b}{1-b}}+1,
\frac{2M}{(1-b)R}\bigg]\nonumber\\&+&
bM ~_2F_1\bigg[\frac{1}{2}\bigg(\sqrt{\frac{9-b}{1-b}}-1\bigg),
\frac{1}{2}\bigg(\sqrt{\frac{9-b}{1-b}}+5\bigg),\sqrt{\frac{9-b}{1-b}}+2,
\frac{2M}{(1-b)R}\bigg]\bigg\}+\mathcal{O}(a^2)\bigg],\label{Time}
\end{eqnarray}
where $T$ is the orbital period and $R$ is the radius of the
circular orbit. The later terms in the right hand side is the
correction by the $a$ and the symmetry breaking scale $\eta$ of
global monopole. From Eq.(\ref{Time}), we find that for fixed $R$
the $\eta$ affects the orbital period $T$ only in the case with
nonzero rotation parameter $a$. In Fig. (3), we present the change
of the corrected term $\Delta T^2=\frac{4\pi^2T^2R^3}{M}-1$ with the
parameter $\eta$ in this spacetime. It is shown that the absolute
value $|\Delta T^2|$ increases with the scale $\eta$ for the system
$SR_+$, but decreases with $\eta$ in the system $SR_-$. Moreover, we
also find that the presence of the global monopole make the orbital
period $T$ increase for a prograde particle (i.e., $a>0$) and
decrease for a retrograde one (i.e., $a<0$).
\begin{figure}[ht]
\begin{center}
\includegraphics[width=7cm]{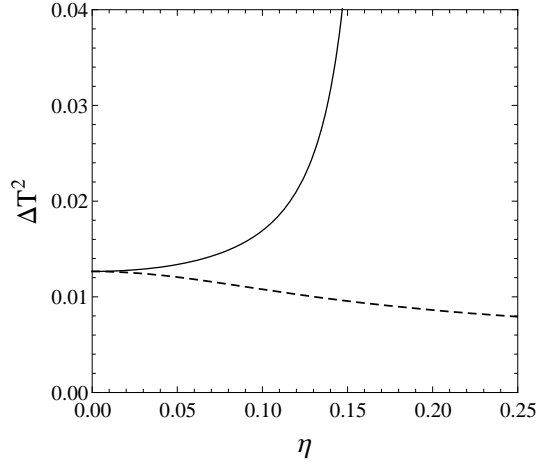}
\caption{The change of the corrected term $\Delta
T^2=\frac{4\pi^2T^2R^3}{M}-1$ in the orbital period $T$ with the
parameter $\eta$.  The solid  and dashed lines are for the systems
$SR_+$ and $SR_-$, respectively. Here we set $M=1$, $a=0.2$ and
$R=10M$.}
\end{center}
\end{figure}

The innermost stable circular orbit (ISCO) of the particle around
the black hole is given by the condition $V_{eff,rr}=0$. For a slow
rotating black hole with the global monopole (\ref{metric3}), we
obtain
\begin{eqnarray}
r_{ISCO}&=&\frac{6M}{1-b}-
\frac{4a\sqrt{2}~3^{-2-\frac{1}{2}\sqrt{\frac{9-b}{1-b}}}}{(1-b)^{5/2}
(\sqrt{\frac{9-b}{1-b}}+1)(\sqrt{\frac{9-b}{1-b}}+2)}\bigg\{9\bigg[\sqrt{(9-b)(10-3b)}
+(3b^2-25b+30)\bigg]\times\nonumber\\&&~_2F_1\bigg[\frac{1}{2}\bigg(\sqrt{\frac{9-b}{1-b}}-3\bigg),
 \frac{1}{2}\bigg(\sqrt{\frac{9-b}{1-b}}+3\bigg),\sqrt{\frac{9-b}{1-b}}+1,
\frac{1}{3}\bigg]+3b\bigg[5\sqrt{(9-b)(1-b)}
+(15-7b)\bigg]\times\nonumber\\&&~_2F_1\bigg[\frac{1}{2}
\bigg(\sqrt{\frac{9-b}{1-b}}-1\bigg), \frac{1}{2}\bigg(\sqrt{\frac{9-b}{1-b}}+5\bigg),\sqrt{\frac{9-b}{1-b}}+2,
\frac{1}{3}\bigg]+b\bigg[\sqrt{(9-b)(1-b)}
+(1+b)\bigg]\times\nonumber\\&&~_2F_1\bigg[\frac{1}{2}
\bigg(\sqrt{\frac{9-b}{1-b}}+1\bigg), \frac{1}{2}\bigg(\sqrt{\frac{9-b}{1-b}}+7\bigg),\sqrt{\frac{9-b}{1-b}}+3,
\frac{1}{3}\bigg]\bigg\}+\mathcal{O}(a^2),\label{risco}
\end{eqnarray}
Obviously, the ISCO radius $r_{ISCO}$ decreases with the rotation
parameter $a$, which is similar to that in the usual Kerr black
hole. In Fig.(4), we set $M=1$ and plotted the variety of the ISCO
radius $r_{ISCO}$ with the parameter $\eta$ in the slowly rotating
black hole spacetime with the global monopole.
\begin{figure}[ht]
\begin{center}
\includegraphics[width=7cm]{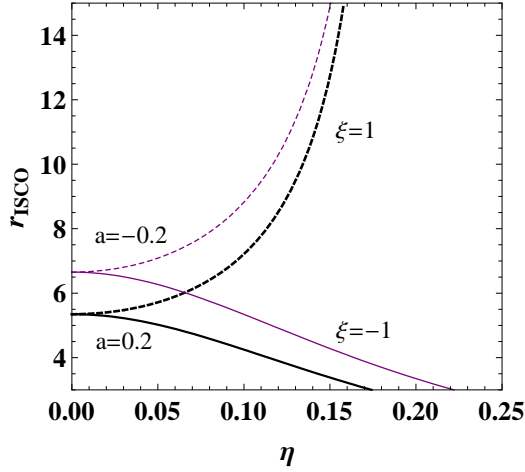}
\caption{The change of the ISCO radius $r_{ISCO}$ with the
parameters $\eta$ and $a$.  The dashed and solid lines are for  the
systems $SR_+$ and $SR_-$, respectively. The thin and thick lines
correspond to the cases with $a=-0.2$ and $a=0.2$. Here we set
$M=1$.}
\end{center}
\end{figure}
It is shown that the ISCO radius $r_{ISCO}$ increases with the scale
$\eta$ for the system $SR_+$, but decreases with $\eta$ in the
system $SR_-$. Moreover,  we find that the ISCO radius in the system
$SR_+$ is larger than that in the system $SR_-$ for fixed $a$.

Let us now compute the effect of the symmetry breaking scale $\eta$
on the radiative efficiency $\epsilon$ in the thin accretion disk
model, which is defined by
\begin{eqnarray}
\epsilon=1-E(r_{ISCO}).
\end{eqnarray}
This quantity corresponds to the maximum fraction of energy being
radiated when a test particle accretes into a central black hole.
For Schwarzschild and extremal Kerr black holes, $\epsilon\sim 0.06$
and $\epsilon\sim 0.42$, respectively. For a slowly rotating black
hole with the global monopole, the radiative efficiency $\epsilon$
can be expressed as
\begin{eqnarray}
\epsilon&=&1-\frac{2\sqrt{2}}{3}\sqrt{1-b}+
\frac{a~3^{-2-\frac{1}{2}\sqrt{\frac{9-b}{1-b}}}}{M
(\sqrt{\frac{9-b}{1-b}}+1)}\bigg\{\sqrt{(9-b)}\bigg[\sqrt{(9-b)}+\sqrt{(1-b)}\bigg]
\times\nonumber\\&&~_2F_1\bigg[\frac{1}{2}\bigg(\sqrt{\frac{9-b}{1-b}}-3\bigg),
 \frac{1}{2}\bigg(\sqrt{\frac{9-b}{1-b}}+3\bigg),\sqrt{\frac{9-b}{1-b}}+1,
\frac{1}{3}\bigg]+\frac{2b}{3}\times\nonumber\\&&~_2F_1\bigg[\frac{1}{2}
\bigg(\sqrt{\frac{9-b}{1-b}}-1\bigg), \frac{1}{2}\bigg(\sqrt{\frac{9-b}{1-b}}+5\bigg),\sqrt{\frac{9-b}{1-b}}+2,
\frac{1}{3}\bigg]\bigg\}+\mathcal{O}(a^2),\label{xlv}
\end{eqnarray}
It is clear that the radiative efficiency $\epsilon$ increases with
the rotation parameter $a$ in these two global monopole black holes.
The effect of the symmetry breaking scale $\eta$ on the radiative
efficiency $\epsilon$ is shown in Fig. (5).
\begin{figure}[ht]
\begin{center}
\includegraphics[width=7cm]{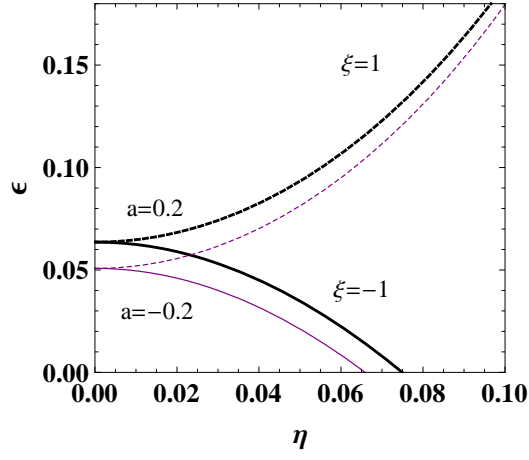}
\caption{The change of the radiative efficiency $\epsilon$ with the
parameter $\eta$ and $a$. The dashed and solid lines are for the
systems $SR_+$ and $SR_-$, respectively. The thin and thick lines
correspond to the cases with $a=-0.2$ and $a=0.2$. Here we set
$M=1$.}
\end{center}
\end{figure}
It tells us that with the scale $\eta$ the radiative efficiency
$\epsilon$ increases for the system $SR_+$, but decreases for the
system $SR_-$. Moreover, we also note that for the phantom black
hole the radiative efficiency $\epsilon$ is positive only for the
case $\eta\leq \eta_c$. This could be explained by a fact that for a
phantom black hole with $\eta\geq \eta_c$, its capability of
capturing particle could become so weak that the accreted matter
around it is very dilute, which could lead to that the radiation can
not be generated because of lacking of the enough stress and dynamic
friction in the accretion disk model. The critical value $\eta_c$
can be approximated as
\begin{eqnarray}
\eta_c=0.0705+0.0227\frac{a}{M}+\mathcal{O}(a^2).
\end{eqnarray}
It means that the critical value $\eta_c$ increases with the
rotation parameter $a$, which is also shown in Fig.(6).
\begin{figure}[ht]
\begin{center}
\includegraphics[width=7cm]{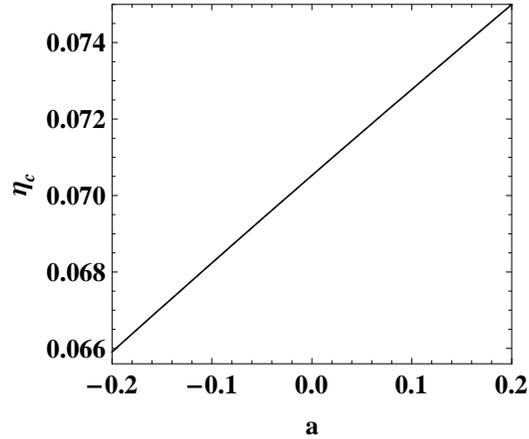}
\caption{The change of the critical value $\eta_c$ with the rotation
parameter $a$ for the system $SR_-$. Here we set $M=1$.}
\end{center}
\end{figure}

\section{summary}

In this paper we present firstly a four-dimensional spherical
symmetric black hole with phantom global monopole and find that the
scale of symmetry breaking $\eta$ affects the radius of the black
hole horizon and a deficit solid angle. For the system $SR_-$, the
solid angle is surplus rather than deficit as in the systems $SR_+$.
Then, we obtain a slowly rotating black hole solution with global
monopole by solving Einstein's field equation. We find that presence
of global monopole makes the metric coefficient $g_{t\phi}$ contain
the hypergeometric function of the polar coordinate $r$, which is
more complex than that in the usual slowly rotating black hole. We
study the property of the angular velocity of the horizon
$\Omega_H$, which is connected with the region where the
super-radiance occurs in the black hole background. With the
increase of $\eta$, the angular velocity $\Omega_H$ first decreases
slowly and then increases rapidly for the system $SR_+$, but
increases monotonically with  $\eta$ for the system $SR_-$.

We also analyze the effects of the scale of symmetry breaking $\eta$
on the Kepler's third law, the innermost stable circular orbit and
the radiative efficiency $\epsilon$ in the thin accretion disk
model. Our results show that only in the case with nonzero rotation
parameter $a$ the orbital period $T$ depends on the scale of
symmetry breaking $\eta$ for fixed orbital radius $R$. The presence
of the global monopole make the orbital period $T$ increase for a
prograde particle (i.e., $a>0$) and decrease for a retrograde one
(i.e., $a<0$). All of the absolute value of the corrected term to
Kepler's Third Law (i.e.,$|\Delta T^2|$), the ISCO radius $r_{ISCO}$
and the radiative efficiency $\epsilon$ in the thin accretion disk
model increase with the scale $\eta$ for the system $SR_+$, but
decrease with $\eta$ in the system $SR_-$. Moreover, we also find
that for the phantom black hole the radiative efficiency $\epsilon$
is positive only for the case $\eta\leq \eta_c$. The threshold value
$\eta_c$ increases with the rotation parameter $a$.

\section{\bf Acknowledgments}
This work was  partially supported by the National Natural Science
Foundation of China under Grant No.11275065, the NCET under Grant
No.10-0165, the PCSIRT under Grant No. IRT0964,  the Hunan
Provincial Natural Science Foundation of China (11JJ7001) and the
construct program of key disciplines in Hunan Province. J. Jing's
work was partially supported by the National Natural Science
Foundation of China under Grant Nos. 11175065, 10935013; 973 Program
Grant No. 2010CB833004.

\appendix
\section{The form of $h(r)$}
Here we present the form of $h(r)$ by solving the radial equation (\ref{radial}).
Defining $z=\frac{2M}{(1-b)r}$, the radial equation (\ref{radial}) can be expressed as
\begin{eqnarray}
z(1-z)\frac{d^2h(z)}{dz^2}+2(1-z)\frac{dh(z)}{dz}+2\bigg[1-\frac{b}{(1-b)z}\bigg]h(z)
=0.\label{radial1}
\end{eqnarray}
Employing the transformation $h(z)=z^{\alpha}(1-z)^{\beta}H(z)$, we
can write Eq. (\ref{radial1}) into the standard form of the
hypergeometric equation
\begin{eqnarray}
z(1-z)\frac{d^2H(z)}{dz^2}+[c-(1+a_1+b_1)z]\frac{dH(z)}{dz}-a_1b_1H(z)=0,
\end{eqnarray}
with
\begin{eqnarray}
c_1=2+2\alpha,~~~~~~~a_1=\alpha+\beta+2,~~~~~~b_1=\alpha+\beta-1.
\end{eqnarray}
Because of the constraint from coefficient of $H(z)$, the
power coefficients $\alpha$ and $\beta$ must satisfy the second-order
algebraic equations
\begin{eqnarray}
\beta=0,~~~~~~~~\alpha(\alpha+1)-\frac{2}{1-b}=0.
\end{eqnarray}
Considering the asymptotical behavior $h(r)$ at spatial infinite
$r\rightarrow\infty$, we choose
$\alpha=\frac{1}{2}[\sqrt{\frac{9-b}{1-b}}-1]$. Then, the function
$h(r)$ has the form
\begin{eqnarray}
h(r)=\bigg[\frac{2M}{(1-b)r}\bigg]^{\frac{1}{2}(\sqrt{\frac{9-b}{1-b}}-1)}
~_2F_1\bigg[\frac{1}{2}\bigg(\sqrt{\frac{9-b}{1-b}}-3\bigg),
 \frac{1}{2}\bigg(\sqrt{\frac{9-b}{1-b}}+3\bigg),\sqrt{\frac{9-b}{1-b}}+1,
\frac{2M}{(1-b)r}\bigg].
\end{eqnarray}


\vspace*{0.2cm}
\begin{thebibliography}{99}

\baselineskip=0.6 cm \baselineskip=0.6 cm

\bibitem{Caldwell} Caldwell  R R 2002 \textit{Phys. Lett. B} {\bf545}, 23 (arXiv: astro-ph/9908168)

\bibitem{ph1} McInnes B 2002 \textit{J. High Energy Phys.} {\bf08} 029 (arXiv: hep-th/0112066)

\bibitem{ph2} Nojiri S and Odintsov S D 2003 \textit{Phys. Lett. B} {\bf562} 147(arXiv:hep-th/0303117)

\bibitem{ph3} Chimento L P and Lazkoz R 2003 \textit{Phys. Rev. Lett.} {\bf91} 211301(arXiv:gr-qc/0307111)

\bibitem{ph4} Boisseau B, Esposito-Farese G, Polarski D and Starobinsky  A A 2000 \textit{Phys. Rev. Lett.} {\bf85}
 2236 (arXiv:gr-qc/0001066)

\bibitem{ph5} Gannouji R, Polarski D, Ranquet A and Starobinsky A A 2006
\textit{J. Cosmol. Astropart. Phys.} {\bf09} 016


\bibitem{ph6} Caldwell R R,  Kamionkowski M and  Weinberg N N 2003
\textit{Phys. Rev. Lett.} {\bf91} 071301

Nesseris S and Perivolaropoulos L 2004 \textit{Phys. Rev. D} {\bf70}
123529 (arXiv:astro-ph/0410309)

Nojiri S and Odintsov S D 2003 \textit{Phys. Lett. B} {\bf571} 1
(arXiv:hep-th/0306212)

Singh P, Sami M and Dadhich N 2003 \textit{Phys. Rev. D} {\bf68}
023522 (arXiv: hep-th/0305110)

Hao J G and Li X Z 2004 \textit{Phys. Rev. D} {\bf70} 043529 (arXiv:
astro-ph/0309746)

Saridakis E N, Gonzalez-Diaz P F and Siguenza C L 2009
\textit{Class. Quant. Grav.} {\bf26} 165003


\bibitem{ph7} Melchiorri A, Mersini-Houghton L, Odman C J and Trodden M 2003 \textit{Phys. Rev. D} {\bf68} 043509
  (arXiv:astro-ph/0211522)

  Ainou M A 2013 \textit{Phys. Rev. D} {\bf87} 024012 (arXiv:1209.5232)


\bibitem{EBab} Babichev E, Dokuchaev V and Eroshenko Y 2004 \textit{Phys. Rev. Lett.} {\bf93} 021102


\bibitem{Sb}  Chen S, Jing J and Pan Q 2009 \textit{Phys. Lett. B} {\bf670} 276

 Chen S, Jing J 2009 \textit{J. High Energy Phys.} {\bf03} 081


\bibitem{bw}  He X, Wang B, Wu S and Lin C 2009 \textit{Phys. Lett. B} {\bf673} 156


\bibitem{Sb1} Chen S and Jing J 2005 \textit{Class. Quant. Grav.} {\bf22} 4651

\bibitem{pbh1} Gibbons G W and Rasheed D A 1996 \textit{Nucl. Phys. B} {\bf476} 515
(arXiv:hep-th/9604177)

Cl'ement G, Fabris J C and Rodrigues M E 2009 \textit{Phys. Rev. D}
 {\bf79} 064021 (arXiv:0901.4543)

Azreg-Ainou M, Cl'ement G,  Fabris J C and Rodrigues M E 2011
\textit{Phys. Rev. D} {\bf83} 124001 (arXiv:1102.4093).

\bibitem{pbh2} Gao C J and Zhang S N 2006 arXiv:hep-th/0604114

\bibitem{pbh3} Bronnikov K A and Fabris J C 2006 \textit{Phys. Rev. Lett.} {\bf96} 251101 (arXiv:gr-qc/0511109).

\bibitem{pbh4} Rodrigues M E and Oporto Z A A 2012 \textit{Phys. Rev. D} {\bf85} 104022(arXiv:1201.5337)

\bibitem{pbh5} Jardim D F,Rodrigues M E and Houndjo M J S 2012 arXiv:1202.2830

\bibitem{pbh6} Nakonieczna A, Rogatko M and Moderski R 2012 \textit{Phys. Rev. D} {\bf86} 044043
(arXiv:1209.1203)

\bibitem{pbh7} Azreg-Ainou M 2012 arXiv:1209.5232

\bibitem{pbh8} Bolokhov S V, Bronnikov K A and Skvortsova M V 2012 \textit{Class. Quant. Grav.} {\bf29},
245006

\bibitem{Barriola} Barriola M and Vilenkin A 1989 \textit{Phys. Rev. Lett.} {\bf63} 341


\bibitem{gb1} Yu H 2002 \textit{Phys. Rev. D} {\bf65} 087502

\bibitem{gb2} Paulo J, Pitelli M and Letelier P 2009 \textit{Phys. Rev. D} {\bf80} 104035

\bibitem{gb3} Chen S and Jing J  \textit{Mod. Phys. Lett. A} {\bf23} 359

\bibitem{gb4} Rahaman F, Ghosh P, Kalam M and Gayen K 2005 \textit{Mod. Phys. Lett.A} {\bf20} 1627.




\end{thebibliography}
\end{document}